\documentclass[twocolumn]{iopart}
\usepackage{hyperref}
\usepackage{graphicx}
\usepackage{color}
\usepackage{ulem}
\usepackage{amssymb}
\usepackage{mathrsfs}
\usepackage{cite}
\usepackage{subfigure}
\usepackage{bm}
\renewcommand{\abstractname}{Abstract}
\renewcommand{\figurename}{Figure}
\usepackage{caption}
\begin{document}

\title{How zonal states beat-driven by Alfv\'en eigenmodes stabilize micro-turbulence in tokamak plasmas} 

\author{Liu Chen$^{1,2}$, Zhiyong Qiu$^{3,2}$, Tiancheng Liu$^1$, Fulvio Zonca$^{2,1}$,   and  Qian Fang$^1$}

\address{$^1$ School of Physics, Zhejiang University, Hangzhou 310027, People's Republic of China}
\address{$^2$ Center for Nonlinear Plasma Science and   C.R. ENEA Frascati, Frascati 00044, Italy}
\address{$^3$   CAS Key Laboratory of Frontier Physics in Controlled Nuclear Fusion, Hefei Institute of Physical Science, Hefei 230031, People's Republic of China}

\address{$\ast$ E-mail of corresponding author:  zqiu@ipp.ac.cn}

\begin{abstract}
The effects of zonal states (ZS) beat-driven by Alfv\'en eigenmodes (AEs)   on micro-turbulence stability are  investigated;  taking toroidal Alfv\'en eigenmode (TAE) and ion temperature gradient mode (ITG) as  an example. The analysis is carried out by deriving the nonlinear gyrokinetic equation describing particle response to ITG,  in  the presence of TAE beat-driven ZS.  It is found that, the phase space zonal structure generated by AEs could lead to a significant reduction in the local  nonuniformity drive and, thereby, stabilize the local ITG instabilities. The zonal flow induced ``shearing", on the other hand, contributes only to a real frequency shift and, typically, negligibly small additional stabilization. 

\end{abstract}

\maketitle
\ioptwocol

High confinement  is crucial for the achievement of high fusion gain and net power generation   in magnetically confined fusion reactors such as  ITER \cite{KTomabechiNF1991} or CFEDR \cite{RDingPST2025}.  For high confinement, two aspects are involved. The first is the bulk thermal  plasma confinement, typically determined by the anomalous transport by micro-scale drift wave (DW) type instabilities driven by   expansion free energy associated with plasma nonuniformities \cite{WHortonRMP1999}. The other is the good  confinement of fusion-born energetic alpha  particles, that powers the reactor and maintains it at the high-temperature required  for the self-sustained burning, and is generally recognised   to be determined by typically meso-scale shear Alfv\'enic wave (SAW) type instabilities \cite{LChenRMP2016,AFasoliNF2007,MSalewskiNF2025}.   These two categories of instabilities, characterized with  different spatial and temporal scales,  and dominating the transport of particles with different energy range, are usually  studied separately. For example, DWs with the wavelength being much smaller than the characteristic orbit width of energetic particles (EPs),  and frequency being much lower than typical EP characteristic frequencies, are believed to be little affected by EPs, and consistently, contribute little to EP transport \cite{WZhangPRL2008,ZFengPoP2013}. As a natural consequence, EPs will not, at least, significantly  affect bulk plasma confinement through the regulation of DW instability.  However, in recent experiments with high power EP heating, bulk plasma confinement are shown to be significantly improved \cite{ADiSienaPRL2021,HHanNature2022,MRomanelliPPCF2010}, and various mechanisms  are proposed to explain the underlying mechanisms \cite{FZoncaPPCF2015,JCitrinPPCF2023,ZQiuPST2025a}, including dilution \cite{GTardiniNF2007}, electromagnetic stabilization \cite{MRomanelliPPCF2010}, and resonant stabilization \cite{ADiSienaNF2018}.   The applicability of these mechanisms to future reactors, of course,   remains to be investigated \cite{ZQiuPST2025a}, due to the different parameter regimes with respect to present day machines \cite{FZoncaNJP2015,MFalessiPoP2019}, such as the temperature and/or density ratios between bulk plasmas and EPs, that determine the relative spatiotemporal scales between DW and SAW   instabilities.

It has also been conjectured  that the zonal states (ZS)  \cite{MFalessiPoP2019,MFalessiNJP2023} generated by EP-driven instabilities may regulate DW via ``shearing" \cite{ADiSienaJPP2021,GBrochardPRL2024}.  This process was investigated in Ref. \citenum{QFangNF2025}, where the local  ITG eigenmode equation in the presence of TAE beat-driven ZS is derived and solved  by  assuming $|\omega|\gg \omega_{d,i}, \omega_{tr,i}$ and    taking the lowest-order nonlinear particle response to ITG.  Here, $\omega$, $\omega_{d,i}$ and $\omega_{tr,i}$ are, respectively, the mode frequency,  the thermal ion magnetic drift frequency and transit frequency.     It was found that,  under the above limit, the TAE beat-driven ZS will have an overall weakly  destabilization effect on ITG stability, contrary to usual expectations.

In recent DIII-D experiments, the improved confinement is observed in the presence of finite amplitude TAE \cite{XDuPRL2025}. It is observed that, the ITG intensity is regulated by TAE,  while   the  change of TAE polarization is shown to be  a necessary condition for ITG regulation;  implying that zonal flow (ZF) generation that changes the TAE polarization  may dominate the observed ITG regulation. It is also observed that, during the same experiment campaign, zonal current (ZC) is generated \cite{WHeidbrinkPRL2026}.

To clarify the above discrepancy  of Ref. \citenum{QFangNF2025}  prediction with experimental observation, in this work, the effects of SAW instability beat-driven ZS on micro-scale DW instability is re-visited, taking TAE and ITG as an example. The analysis is carried out  by considering the free energy source from the  renormalized   diamagnetic drift frequency in the presence of TAE and the beat-driven ZS. The obtained nonlinear ITG eigenmode equation is found to  have the same structure  as that of linear ITG, with only the diamagnetic drift frequency being modified, which thus, contains all nonlinear effects due to TAE-ITG cross scale couplings mediated by ZS, and gives results with high certainty.  It is found that, the TAE-induced phase space zonal structure (PSZS) could  significantly  reduce the ITG instability drive, while the ZF induced ``shearing", will only contribute to a nonlinear frequency shift, and, typically, negligible effects on  the ITG stability.  The  analysis of Ref. \citenum{QFangNF2025} is then re-visited (see Appendix), and the  origin of the previous discrepancy with experimental observations is clarified and resolved.

For the clarity of the derivation, in this work, a relatively simple limit is taken, i.e.,  we consider the ITG is electrostatic, the ZS is  beat-driven by TAE, and the particle response to ZS is neoclassical.    We note that, the ZS  beat-driven   by SAW instabilities, was originally formulated  in Refs. \citenum{LChenNF2025}, and the essential derivations are repeated here only  for the completeness  of the paper.    Note that neoclassical response is  assumed in Refs. \citenum{LChenNF2025}  by taking $|\omega_Z|\simeq 2|\gamma_L|\ll |\omega_{tr,i}|$ and $\gamma_L\ll\omega_r$, so  the obtained results is applicable as the SAW instability growth rate is much smaller than its real frequency;  expected  for  reactor relevant parameter regimes \cite{LChenRMP2016,MSalewskiNF2025,ZQiuPST2025b}.  Here, $\omega_Z$ is the ZS frequency, and $\gamma_{L}$ and $\omega_{0r}$ are, respectively,  the linear growth rate and real frequency of SAW instability.   This theory  is then extended to  various limits  in Ref. \citenum{LChen2026a}, including DIII-D experiments with $|\gamma_L/\omega_{0r}|\sim O(10^{-1})$, and is confirmed by GTC simulation.     In this work, we,  however,   adopt the weakly driven limit  for the clarity of the theory. The extension to the parameter regimes of present machines should be straightforward based on the arguments and discussions  of  Ref. \citenum{LChen2026a}.   The self-consistent evolution of the ZS in the presence of a prescribed    symmetry breaking fluctuation spectrum is also given in Ref. \citenum{MFalessiNJP2023} for general tokamak geometry and arbitrary strength of  fluctuation driven by free energy sources.

To describe the nonlinear  beat-driven of ZS by TAE, and the consequences on ITG stability, nonlinear gyrokinetic theory is adopted. In the gyrokinetic theoretical framework, the   particle distribution function can be written as
\begin{eqnarray}
f=F_0+\frac{q}{m}\delta\phi\partial_E F_0 +\exp(-\mathbb{\rho}\cdot\nabla) \delta H_k,
\end{eqnarray}
with the non-adiabatic particle response $\delta H_k$ derived from nonlinear gyrokinetic equation \cite{EFriemanPoF1982}:
\begin{eqnarray}
\left(\partial_t +v_{\parallel}\partial_l + i\omega_d\right)\delta H_k &=& i \frac{q}{m} \left(\omega\partial_E + \hat{\omega}_*\right) F_0 J_k\delta L_k\nonumber\\
&=& -\sum_k \Lambda^{k'}_{k''} J_{k'}\delta L_{k'}\delta H_{k''}.\label{eq:GKE}
\end{eqnarray}
Here, $F_0$ is the equilibrium distribution function,   $\partial_E F_0=-(m/T)F_0$ for Maxwellian distribution, $\exp(-\mathbb{\rho}\cdot\nabla)$ corresponds to guiding-centre transformation with $\mathbb{\rho}\equiv \mathbf{b}\times \mathbf{v}/\Omega_c $,  $\Omega_c\equiv B_0q/(mc)$ is the gyrofrequency,   $\partial_l$ represents the derivative along the equilibrium magnetic field line, $\omega_d=(v^2_{\perp}+2v^2_{\parallel})/(2 \Omega_c R)(k_r\sin\theta+k_{\theta}\cos\theta)$ is the magnetic drift frequency, 
 $\hat{\omega}_*\equiv \mathbf{k}\cdot \mathbf{b} \times \nabla F_0/\Omega_c$, and $\omega_*\equiv (T/m)\hat{\omega}_*$ is the corresponding  diamagnetic drift frequency associated with plasma nonuniformity.   Furthermore, $J_k\equiv J_0(k_{\perp}\rho)$ is the Bessel function of zero index denoting finite Larmor radius effects, $\delta L_k\equiv \delta\phi_k-v_{\parallel}\delta A_{\parallel k}/c$  is  the scalar potential in the guiding-centre frame, $\delta\phi$ is the scalar potential, $\delta A_{\parallel}$ is the parallel component of the vector potential related to the perturbed magnetic field, and $\Lambda^{k'}_{k''}\equiv (c/B_0) \mathbf{b}\cdot\mathbf{k''}\times \mathbf{k'}$.   In the analysis of SAW instabilities, an alternative field variable $\delta\psi\equiv \omega \delta A_{\parallel}/(c k_{\parallel})$ is introduced for convenience, with  $\delta\phi=\delta\psi$ corresponding to ideal MHD constraint. 
 
For typically  low-$\beta$ plasma of interest to magnetically confined fusion research, the governing equations can be derived from quasi-neutrality condition
\begin{eqnarray}
\frac{n_0 e^2}{T_i}\left(1+\frac{T_i}{T_e}\right)\delta\phi_k=\sum_{s=e,i}\left\langle q_sJ_0\delta H_s\right\rangle_s,\label{eq:qn}
\end{eqnarray}
 and the parallel component of the Amp\'ere's law
 \begin{eqnarray}
 \nabla^2_{\perp}\delta A_{\parallel k}=\frac{4\pi}{c}\sum_s\left\langle q_s v_{\parallel s}\delta H_s\right\rangle_k.\label{eq:ampere}
 \end{eqnarray}
 Note that, by writing Eqs. (\ref{eq:qn}) and (\ref{eq:ampere}) in the present form, we assume typical ordering of reactor relevant plasmas \cite{LChenRMP2016}, where dilution effects due to EP \cite{GTardiniNF2007} as well as ``resonant" stabilization  \cite{ADiSienaNF2018} are typically negligible, as demonstrated theoretically  \cite{LChenRMP2016} and confirmed in recent numerical simulations \cite{RIvannovPPCF2026}.   We, however, try to minimize the derivations, and focus instead on the physics picture.

We, first, briefly recapitulate the beat-driven process. 
The linear particle response to TAE $\Omega_0(\omega_0, \mathbf{k}_0)$ can be derived from the linear gyrokinetic equation, which will be used to derive the nonlinear particle responses to zonal structure.  Noting the $k_{\parallel}v_e\gg\omega_0\sim V_A/(2qR_0) \gg k_{\parallel}v_i, \omega_d$ ordering, as well as $\delta\phi_0\simeq \delta\psi_0$  (i.e., $\delta E_{\parallel0}\simeq 0$),  the particle responses to TAE can be derived as
\begin{eqnarray}
\delta H_{0i}&\simeq& \frac{q}{T} \left(1-\frac{\omega_{*i}}{\omega}\right)_0 F_0 J_0\delta\phi_0,\nonumber\\
\delta H_{0e}&\simeq& \frac{q}{T}\left(1-\frac{\omega_{*e}}{\omega}\right)_0F_0\delta\psi_0. \nonumber
\end{eqnarray}

The particle responses to TAE can be substituted into the nonlinear gyrokinetic equation  to derive the particle responses of ZS. Here,  as noted above, we take the ZS to be slowly evolving;  i.e.,       $\omega_Z\ll\omega_{tr,i}$.   The ion response to ZS, can be derived by taking $\delta H_Z=e^{i\lambda_Z}\delta H_{dZ}$, with $\lambda_Z$ obtained from $\omega_{tr}\partial_{\theta}\lambda_Z + \omega_d=0$, and one obtains
\begin{eqnarray}
\left(\partial_t + \omega_{tr}\partial_{\theta}\right)\delta H_{dZ}&=&-i\omega \frac{q}{T}F_0 e^{-i\lambda_Z}\left(\delta\phi_Z-v_{\parallel}\delta A_{\parallel Z}/c\right)\nonumber\\
&&-e^{-i\lambda_Z}\Lambda^{k'}_{k''} J_{k'}\delta L_{k'}\delta H_{k''}.\label{eq:ZS_NL}
\end{eqnarray}

Separating  $\delta H_Z=\overline{\delta H_Z} +\widetilde{\delta H_Z}$, with $\overline{\cdots}$ denoting  transit/bounce averaging,  one can separate equation (\ref{eq:ZS_NL}) into a.c. and d.c. parts. 
We then have, $|\widetilde{\delta H_Z}/\overline{\delta H_Z}|\sim |\omega_Z/\omega_{tr}|\ll1$. Considering only the beat-driven generation of ZS due to TAE self-interaction, one obtains,
\begin{eqnarray}
\delta H_Z=\delta H^{(1)}_Z+\delta H^{(2)}_Z, \nonumber
\end{eqnarray}
with
\begin{eqnarray}
 \delta H^{(1)}_{Zi}&\simeq& \frac{q}{T}F_0 \left|\overline{e^{i\lambda_Z}}\right|^2 \left(\delta\phi_Z - \sigma \overline{v}_{\parallel}\delta A_{\parallel Z}/c\right), \label{eq:Zi_lin}\\
 \delta H^{(2)}_{Zi}&\simeq& \frac{c}{B_0}k_{\theta 0} \frac{q}{T}F_0\frac{\omega_{*0i}}{\omega^2_0} \left| \overline{e^{i\lambda_Z}}\right|^2  \partial_r \overline{\left|\delta \phi_0\right|^2}.
\end{eqnarray}
Here, $\sigma=1$ for the circulating ions and $\sigma=0$ for trapped ions.

The electron response to ZS, can be derived similarly, and one obtains
\begin{eqnarray}
\delta H^{(1)}_{Ze}&\simeq& \frac{q}{T} F_0\left(\delta\phi_Z-\sigma\overline{v}_{\parallel}\delta A_{\parallel Z}/c\right),\\
\delta H^{(2)}_{Ze}&\simeq& -\frac{c}{B_0}k_{\theta0}\frac{q}{T}F_0\left(\frac{\omega_{*e}}{\omega^2_0} - \frac{k_{\parallel}\overline{v}_{\parallel}}{\omega^2_0}\right)\partial_r \overline{\left|\delta \phi_0\right|^2}. 
\end{eqnarray}

The nonlinear generation of zonal flow  and zonal magnetic field, can thus be obtained, by substituting the particle response into quasi-neutrality condition and parallel  Amp\'ere's law. One then has, for TAE \cite{LChen2026a}
\begin{eqnarray}
\delta\phi_Z&\simeq& \frac{c}{B_0} \frac{1}{\omega^2_{0r}}\left(1+ \eta_i\right) \frac{\partial}{\partial r}   k_{\theta0} \omega_{*in} \left(\Phi^{(2)}_a+\Phi^{(2)}_b\right), \label{eq:ZF}\\
\delta A_{\parallel Z}&\simeq& \frac{c^2}{B_0}\frac{1}{\omega^2_{0r}} \frac{\partial}{\partial r}   k_{\theta 0}k_{\parallel 0}\Phi^{(2)}_a; \label{eq:ZC}
\end{eqnarray} 
here, in terms of the poloidal harmonics, $\Phi^{(2)}_a=\sum_m \Phi^2_m$ and $\Phi^{(2)}_b\simeq \sum_m\Phi_m(\Phi_{m+1}+\Phi_{m-1})$. 
We note that,  equations (\ref{eq:ZF}) and (\ref{eq:ZC}) were originally derived in Ref. \citenum{LChenNF2025} for ZS generation by reversed shear Alfv\'en eigenmode (RSAE), while the expressions can be generally applied to other SAW instabilities, noting their respective linear properties.

For the effect of ZS on ITG stability of interest here, the nonlinear gyrokinetic equation for ITG shall be derived, based on which, we show the instability drive from plasma nonuniformity could be  significantly reduced in the presence of AE driven ZS;  in particular,  the PSZS $\delta H^{(2)}_{Zi}$ plays a crucial role. 

Considering the electrostatic limit  and neglecting    trapped ion  effects, nonlinear  thermal ion response to ITG can be derived from equation (\ref{eq:GKE})
\begin{eqnarray}
&&\left(\partial_t+v_{\parallel}\partial_l + i \omega_d\right)\delta H_I = -i\frac{q}{T} F_0 (\omega-\omega_*) J_I\delta\phi_I\nonumber\\
&&-\frac{c}{B_0}  \mathbf{b} \cdot\mathbf{k}_Z\times\mathbf{k}_I \left[J_I\delta\phi_I\delta H_Z - J_Z\delta L_Z\delta H_I\right].\label{eq:GKE_ITG_original}
\end{eqnarray}
Note that equation (\ref{eq:GKE_ITG_original}) is equivalent to equation (18) of Ref. \citenum{QFangNF2025}, and could be rewritten, with $\partial_t=-i\omega$, as  
\begin{eqnarray}
&&\left[-i\left(\omega-\hat{\omega}_Z-i\omega_d\right) + v_{\parallel}\partial_l + i v_{\parallel} \delta k_{\parallel Z}\right]\delta H_I\nonumber\\
&=& -i\frac{q}{T} F_0(\omega-\omega_*) J_I\delta\phi_I - \frac{c}{B_0}  \mathbf{b} \cdot\mathbf{k}_Z\times\mathbf{k}_I J_I\delta\phi_I\delta H_Z;\nonumber\\ \label{eq:ITG_mod}
\end{eqnarray}
here, noting $J_Z\simeq 1$, $\hat{\omega}_Z=(c/B_0)k_{\theta I}\partial_r \delta \phi_Z$ and $\delta k_{\parallel Z}= (k_{\theta I}/B_0) \partial_r\delta A_{\parallel Z}$. From equation (\ref{eq:ZC}) and noting that, for typical tokamak parameters, $k_Z\sim 2k_{\perp0}$, $k_{\perp0}\rho_h\sim O(1)$, $qR/\rho_h\sim O(10^3)$, $\rho_i/\rho_h\sim O(10^{-1})$, $|\delta B_r/B_0|^2\sim O(10^{-7} - 10^{-6})$, and $k_{\theta I}\rho_I\sim O(10^{-1})$ for ITG, we then estimate $|\delta B_Z/B_0|\sim O(10^{-4}- 10^{-3})$ and $|\delta k_{\parallel Z}qR|\sim O(10^{-1}-1)$. Furthermore, noting that for the toroidal ITG mode, one typically has $|\omega|>|v_{\parallel}\partial_l| \geq |v_{\parallel}\delta k_{\parallel Z}|$ \cite{CHsuNF2026} and furthermore, as can be readily observed, the $v_{\parallel}\delta k_{\parallel Z}$ term introduces a  symmetry  breaking along $l(\theta)$ with respect to $\theta=0$.  Consequently, the  $v_{\parallel}\delta k_{\parallel Z}$ term due to $\delta A_{\parallel Z}$ (zonal current) in the left hand side of equation (\ref{eq:ITG_mod}) is negligible in the lowest order. Similarly, the contribution to $\delta H_Z$ due to the $\overline{v}_{\parallel}\delta A_{\parallel Z}$ in equation (\ref{eq:Zi_lin})  is also negligible, as it can be recognized by inspection, comparing Eqs. (\ref{eq:ZF}) and (\ref{eq:ZC}) with the present frequency and wavelength ordering.  Introducing $\omega'=\omega-\hat{\omega}_Z$, equation (\ref{eq:ITG_mod}) then reduces to 
\begin{eqnarray}
&&-i\left(\omega'-k_{\parallel}v_{\parallel} -\omega_d\right)\delta H_I\nonumber\\
&=& -i \frac{q}{T} J_I\delta\phi_I\left[\omega' F_0-\frac{c}{B_0}\frac{T}{q} k_{\theta I}\partial_r\left(F_0 + \delta H^{(2)}_Z\right)\right.\nonumber\\
&&\left. -\frac{c}{B_0} k_{\theta I} F_0 \left(\left|e^{i\lambda_Z}\right|^2 - 1\right)\partial_r J_Z\delta \phi_Z\right].\label{eq:GKE_ITG}
\end{eqnarray}
Here, $k_{\parallel}=-i\partial_l$ and we have applied equation (\ref{eq:Zi_lin}).

Noting the expression of $\delta H^{(2)}_{Zi}$ and that ZS varies faster than $F_0$ in the radial direction,  one has
\begin{eqnarray}
\partial_r\left(F_0 +\delta H^{(2)}_Z\right)\simeq (1+\alpha_Z) \partial_r F_0,
\end{eqnarray}
with 
\begin{eqnarray}
\alpha_Z\equiv \frac{c^2}{B^2_0}k^2_{\theta 0}\frac{1}{\omega^2_{0r}}   \left|e^{i\lambda_Z}\right|^2  \partial^2_r \overline{|\delta\phi_0|^2}.
\end{eqnarray}
For $\delta\phi_0$ localized around $r_0$, we have, $ \partial^2_r \overline{|\delta\phi_0|^2}<0$  and, thus, $\alpha_Z<0$ and   the $\delta H^{(2)}_{Zi}$ term  plays  a stabilizing role  on ITG near the peak of TAE intensity.  \footnote{It is noteworthy that,  $\alpha_Z$ may change sign for the ITG with their mode structure having a misalignment with the peak of TAE intensity. In that case, the ITG drive may be enhanced by TAE induced PSZS, though the enhancement is weaker than the reduction at $r_0$. The effect can, however, still be quite significant in present-day machines with strong EP drive. }
 The $\delta\phi_Z$ (ZF) induced ``shearing" effects, meanwhile,  correspond to the last term on the right hand side of equation (\ref{eq:GKE_ITG}). Noting that $|\lambda_Z|\sim |k_Z\rho_{di}|<1$, it then gives rise to, noting equation (\ref{eq:ZF}) for $\delta\phi_Z$,  $(c/B_0) k_{\theta I} F_0 \rho^2_{di}\partial^3_r\delta\phi_Z>0$; i.e., stabilization.  Quantitatively, since $|k_Z|\sim 1/\rho_h$, its contribution is, typically, $O(|\rho_i/\rho_h|^2\sim T_i/T_h\sim 10^{-2})$ smaller than that due to $\delta H^{(2)}_{Zi}$ generated, directly by radial corrugations due to AE fluctuations.

Neglecting, for now,  trapped-electron effects, the electron response is, thus, adiabatic. The corresponding electrostatic ITG eigenmode equation in the presence of TAE beat-driven ZS can then be derived from the quasi-neutrality condition and formally written as 
\begin{eqnarray}
  \left[1+\frac{1}{\tau}    -  \left\langle J^2_I\frac{1}{n_0}  \frac{\omega'-\omega'_*}{\omega'-k_{\parallel}v_{\parallel} -\omega_d} F_0 \right\rangle \right]\delta\phi_I=0. \label{eq:ITG_DR}
\end{eqnarray}
Here, $\tau=T_e/T_i$,  $\omega'=\omega -\hat{\omega}_Z$,  
 $\langle\cdots\rangle$ denotes velocity integral,  
\begin{eqnarray}
\omega'_*\equiv (1+\alpha_Z)\omega_*, 
\end{eqnarray} 
and $\delta\phi_I$ can usually be cast  in the ballooning-mode representation.

 Equation (\ref{eq:ITG_DR}) has exactly  the same structure as the well-known  linear eigenmode equation  of  ITG in tokamaks \cite{CHsuNF2026}, with the frequency $\omega'$ having a real frequency shift $\hat{\omega}_Z$, while the diamagnetic frequency $\omega'_*$ is reduced, noting $\alpha_Z<0$,   by the factor $(1+\alpha_Z)$. Thus, we can conclude that the TAE beat-driven ZS, will significantly stabilise   ITG by reducing the effective free energy from the plasma nonuniformities, with the dominant contribution from PSZS induced by the TAE.  On the other hand, ZF will only introduce a frequency shift, while, typically,  having negligible effect on ITG stability due to the spatiotemporal scale separation with TAE.     This can be more clearly identified, as  equation (\ref{eq:ITG_DR}), in the $\eta_i\gg1$ limit,  gives   $\omega'^3\simeq -k^2_{\parallel}c^2_S \omega'_{*Ti}/2$ or $\omega'^2\simeq - \omega'_{*Te}\omega_d$ for the slab and toroidal branches,  respectively, with $\omega_{*Ti}$ being the diamagnetic drift frequency associated with temperature nonuniformity. Thus the TAE induced PSZS will stabilize both branches  through the reduced drive.  More quantitatively, we can establish an AE intensity threshold ``sufficient" for suppressing ITG by letting $\omega'_*=0$ or $1+\alpha_Z\simeq0$; leading to 
 \begin{eqnarray}
 \left| \frac{\partial^2}{\partial r^2} \left|\frac{\delta B_{0r}}{k_{\parallel0}B_0}\right|^2\right|_{th}\simeq 1. \label{eq:threshold}
 \end{eqnarray}
 Taking $|\partial/\partial_r|\simeq |k_Z|\simeq 2|k_{\theta0}|\simeq 2/\rho_h$ and $|k_{\parallel0}|\simeq 1/(2qR)$, equation (\ref{eq:threshold}) then yields the following estimate
 \begin{eqnarray}
 \left|\frac{\delta B_{0r}}{B_0}\right|^2_{th} \simeq \frac{1}{16} \left|\frac{\rho_h}{qR}\right|^2\sim O(10^{-7} - 10^{-6}),
 \end{eqnarray}
 where we have taken $qR/\rho_h\sim O(10^{-3})$. Note that this estimated threshold is reasonably consistent with experimental measurements.

As shown in details in the  \ref{sec:appendix}, the difference  between the current analysis and that of Ref. \citenum{QFangNF2025} is due to   Ref. \citenum{QFangNF2025}   neglecting  order $|k_{\parallel}v_{\parallel}/\omega|$  and $|\omega_{d,i}/\omega|$ terms in the nonlinear ion response.

In summary, the effects of TAE beat-driven ZS on ITG stability is analyzed here in order to understand the behavior  of bulk ion confinement in the presence of energetic particles. It is found that,  while the ITG could be significantly stabilized by the PSZS-induced reduction in the instability drive,  
 the ZF-induced ``shearing" effect,  typically, contributes negligibly  to ITG stability; clarifying, thus,  a long-standing misunderstanding of this issue.  

It is noteworthy that, while a simplified scenario is considered, i.e., ITG is predominantly electrostatic, and only circulating ion contribution to ITG  is considered, the approach of deriving the renormalised  nonlinear gyrokinetic equation in the presence of TAE-induced ZS is general, and can be applicable to other types of AEs and micro-instabilities; e.g., the trapped electron mode.    These topics will be investigated in the future. Finally, we note that the present analysis was carried out assuming typical parameter orderings that apply in reactor relevant fusion plasmas, where EP induced dilution and/or ``resonant" effects are expected to be negligible. We also note that the expression of thermal ion stabilizing PSZS is derived under fairly general assumptions of shear Alfv\'en waves (SAWs) with ideal $\delta E_{\parallel}\simeq 0$.  It, thus, suggests that the same stabilizing effect could be achieved via SAWs generated by external antennas either directly or via wave beating. This conjecture could be tested either in experiments or simulations.

\section*{Acknowledgement}
This work was supported by the National Key R\&D Program of China (Nos. 2024YFE03230500 and 2024YFE030170000),     the Strategic Priority Research Program of Chinese Academy of Sciences (No. XDB0790000),   the National Natural Science
Foundation of China (Grant Nos. 12275236 and  12625513),  and  the  International Partnership Program of Chinese Academy of Sciences (Grant No. 145GJHZ2025017BS).   This work has also been carried out within the framework of the EUROfusion Consortium, funded by the European Union via the Euratom Research and Training Programme (Grant Agreement No. 101052200 -
EUROfusion).

\appendix
\section{Comparison with Ref. [Fang et al., Nucl. Fusion \bf{65} (2025) 066004]}
\label{sec:appendix}

In the Appendix, we will show that, the difference  of the present work with respect to Ref. \citenum{QFangNF2025} is caused by the  derivation of formally higher order terms in  $\delta H^{(2)}_{Ii}$, which, however, dominate the overall nonlinear response. 

The nonlinear ion response to ITG $\delta H^{(2)}_{Ii}$ is derived from equation (18) of Ref. \citenum{QFangNF2025}, where $\omega_I\gg k_{\parallel}v_{\parallel}, \omega_d$ is assumed to simplify the derivation.  Keeping the full propagator, the ion response to ITG can be written as
\begin{eqnarray}
\delta H^{(2)}_{Ii} \simeq  \frac{q}{T_i} F_0 J_I\delta\phi_I \mathscr{C}&\times& \left[\frac{|e^{i\lambda_Z}|^2(\omega_{*pi}-\omega_{*i})}{\omega-k_{\parallel}v_{\parallel}-\omega_d} \right.\nonumber\\
&&\left.- \frac{\omega_{*pi}(\omega-\omega_{*i})}{(\omega-k_{\parallel}v_{\parallel}-\omega_d)^2} \right],\label{eq:ITG_NL_app}
\end{eqnarray}
with $\mathscr{C}\equiv (c^2/B^2_0)(k^2_{\theta0}/\omega^2_{0r})\partial^2_r|\delta\phi_0|^2$. 

Noting the expression of $\delta H^{(1)}_{Ii}$,  equation (\ref{eq:ITG_NL_app}) can be re-written as 
\begin{eqnarray}
\delta H^{(2)}_{Ii}&\simeq& \omega_{*pi} \mathscr{C}  \frac{\partial\delta H^{(1)}_{Ii}}{\partial\omega}\nonumber\\
& +& \frac{q}{T_i} F_0J_I\delta\phi_I\mathscr{C} \frac{(|e^{i\lambda_Z}|^2-1)\omega_{*pi} - |e^{i\lambda_Z}|^2 \omega_{*i}}{\omega-k_{\parallel}v_{\parallel}-\omega_d},  \nonumber
\end{eqnarray}
with the first term on the right hand side of the equation  related to the zonal flow induced ``shearing", while the second term is related to the modification of the equilibrium by  PSZS.

Neglecting the $(|e^{i\lambda_Z}|^2-1)$ term by noting the  $k_Z\rho_i\sim \sqrt{T_i/T_h}\ll1$ ordering, assuming adiabatic electron response to ITG, and substitute particle response into quasi-neutrality condition, we have
\begin{eqnarray}
\mathscr{E}_I(\omega) +\omega_{*pi} \mathscr{C}\partial_{\omega}\mathscr{E}_I + \tau\mathscr{C} \left\langle J^2_I \frac{F_0}{n_0} \frac{ |e^{i\lambda_Z}|^2 \omega_{*i}} {\omega-k_{\parallel}v_{\parallel}-\omega_d} \right\rangle = 0,\nonumber\\
\end{eqnarray}
with $\tau\equiv T_e/T_i$ being the electron-to-ion temperature ratio, and
\begin{eqnarray}
\mathscr{E}_I(\omega) = 1+\tau-\tau\left\langle J^2_I \frac{F_0}{n_0} \frac{\omega-\omega_{*i}} {\omega-k_{\parallel}v_{\parallel}-\omega_d} \right\rangle 
\end{eqnarray}
being the linear ITG dispersion relation.  Taking $\omega_0$ being the linear eigenmode frequency satisfying $\mathscr{E}_I(\omega_0)=0$, and introducing $\omega=\omega_0+\delta\omega$, with $\delta\omega$ being the ZS induced frequency shift, and one obtains
\begin{eqnarray}
\delta\omega\simeq -\omega_{*pi}\mathscr{C} - \frac{\tau\mathscr{C}}{\partial_{\omega_0}\mathscr{E}_I} \left\langle J^2_I \frac{F_0}{n_0} \frac{ |e^{i\lambda_Z}|^2 \omega_{*i}} {\omega-k_{\parallel}v_{\parallel}-\omega_d} \right\rangle.
\end{eqnarray}

It is evident that, the first term on the right hand side, related to the zonal flow induced shearing, causes a real frequency shift. The second term due to the contribution from PSZS, on the other hand, may have an impact on the ITG stability, consistent with the analysis below equation (\ref{eq:ITG_DR}) of the main text.

\section*{References}
\providecommand{\newblock}{}


\begin{thebibliography}{10}
\expandafter\ifx\csname url\endcsname\relax
  \def\url#1{{\tt #1}}\fi
\expandafter\ifx\csname urlprefix\endcsname\relax\def\urlprefix{URL }\fi
\providecommand{\eprint}[2][]{\url{#2}}

\bibitem{KTomabechiNF1991}
Tomabechi K, Gilleland J, Sokolov Y, Toschi R and Team I 1991 {\em Nuclear
  Fusion\/} {\bf 31} 1135

\bibitem{RDingPST2025}
DING R, CHAN V~S and LI J 2025 {\em Plasma Science and Technology\/} {\bf 27}
  100101

\bibitem{WHortonRMP1999}
Horton W 1999 {\em Rev. Mod. Phys.\/} {\bf 71}(3) 735--778

\bibitem{LChenRMP2016}
Chen L and Zonca F 2016 {\em Review of Modern Physics\/} {\bf 88} 015008

\bibitem{AFasoliNF2007}
Fasoli A, Gormenzano C, Berk H, Breizman B, Briguglio S, Darrow D, Gorelenkov
  N, Heidbrink W, Jaun A, Konovalov S, Nazikian R, Noterdaeme J~M, Sharapov S,
  Shinohara K, Testa D, Tobita K, Todo Y, Vlad G and Zonca F 2007 {\em Nuclear
  Fusion\/} {\bf 47} S264

\bibitem{MSalewskiNF2025}
Salewski M, Spong D, Aleynikov P, Bilato R, Breizman B, Briguglio S, Cai H,
  Chen L, Chen W, Duarte V, Dumont R, Falessi M, Fitzgerald M, Fredrickson E,
  García-Muñoz M, Gorelenkov N, Hayward-Schneider T, Heidbrink W, Hole M,
  Kazakov Y, Kiptily V, Könies A, Kurki-Suonio T, Lauber P, Lazerson S, Lin Z,
  Mishchenko A, Moseev D, Muscatello C, Nocente M, Podestà M, Polevoi A,
  Schneider M, Sharapov S, Snicker A, Todo Y, Qiu Z, Vlad G, Wang X, Zarzoso D,
  Van~Zeeland M, Zonca F and Pinches S 2025 {\em Nuclear Fusion\/} {\bf 65}
  043002

\bibitem{WZhangPRL2008}
Zhang W, Lin Z and Chen L 2008 {\em Phys. Rev. Lett.\/} {\bf 101} 095001

\bibitem{ZFengPoP2013}
Feng Z, Qiu Z and Sheng Z 2013 {\em Physics of Plasmas\/} {\bf 20} 122309

\bibitem{ADiSienaPRL2021}
Di~Siena A, Bilato R, G{\"o}rler T, Navarro A~B, Poli E, Bobkov V, Jarema D,
  Fable E, Angioni C, Kazakov Y~O {\em et~al.\/} 2021 {\em Physical Review
  Letters\/} {\bf 127} 025002

\bibitem{HHanNature2022}
Han H, Park S, Sung C, Kang J, Lee Y, Chung J, Hahm T~S, Kim B, Park J~K, Bak J
  {\em et~al.\/} 2022 {\em Nature\/} {\bf 609} 269--275

\bibitem{MRomanelliPPCF2010}
Romanelli M, Zocco A, Crisanti F and Contributors J~E 2010 {\em Plasma Physics
  and Controlled Fusion\/} {\bf 52} 045007

\bibitem{FZoncaPPCF2015}
Zonca F, Chen L, Briguglio S, Fogaccia G, Milovanov A~V, Qiu Z, Vlad G and Wang
  X 2015 {\em Plasma Physics and Controlled Fusion\/} {\bf 57} 014024

\bibitem{JCitrinPPCF2023}
Citrin J and Mantica P 2023 {\em Plasma Physics and Controlled Fusion\/} {\bf
  65} 033001

\bibitem{ZQiuPST2025a}
QIU Z, CHEN L, ZONCA F, FALESSI M~V, FANG Q, WEI G and CHEN N 2025 {\em Plasma
  Science and Technology\/} {\bf 27} 095101

\bibitem{GTardiniNF2007}
Tardini G, Hobirk J, Igochine V, Maggi C, Martin P, McCune D, Peeters A, Sips
  A, St{\"a}bler A, Stober J {\em et~al.\/} 2007 {\em Nuclear fusion\/} {\bf
  47} 280

\bibitem{ADiSienaNF2018}
Di~Siena A, Görler T, Doerk H, Poli E and Bilato R 2018 {\em Nuclear Fusion\/}
  {\bf 58} 054002

\bibitem{FZoncaNJP2015}
Zonca F, Chen L, Briguglio S, Fogaccia G, Vlad G and Wang X 2015 {\em New
  Journal of Physics\/} {\bf 17} 013052

\bibitem{MFalessiPoP2019}
Falessi M~V and Zonca F 2019 {\em Physics of Plasmas\/} {\bf 26} 022305

\bibitem{MFalessiNJP2023}
Falessi M~V, Chen L, Qiu Z and Zonca F 2023 {\em New Journal of Physics\/} {\bf
  25} 123035

\bibitem{ADiSienaJPP2021}
Di~Siena A, Gorler T, Poli E, Banon~Navarro A, Biancalani A, Bilato R, Bonanomi
  N, Novikau I, Vannini F and Jenko F 2021 {\em Journal of Plasma Physics\/}
  {\bf 87} 555870201

\bibitem{GBrochardPRL2024}
Brochard G, Liu C, Wei X, Heidbrink W, Lin Z, Gorelenkov N, Chrystal C, Du X,
  Bao J, Polevoi A~R, Schneider M, Kim S~H, Pinches S~D, Liu P, Nicolau J~H and
  L\"utjens H 2024 {\em Phys. Rev. Lett.\/} {\bf 132}(7) 075101

\bibitem{QFangNF2025}
Fang Q, Wei G, Chen N, Chen L, Zonca F and Qiu Z 2025 {\em Nuclear Fusion\/}
  {\bf 65} 066004

\bibitem{XDuPRL2025}
Du X~D, Heidbrink W~W, Yan Z, Diamond P~H, McKee G~R, Schmitz L, Van~Zeeland
  M~A, Wang H~Q, Austin M~E, Liu L, Callahan K~J and Shi N 2025 {\em Phys. Rev.
  Lett.\/} {\bf 135}(26) 265101

\bibitem{WHeidbrinkPRL2026}
Heidbrink W~W, Du X~D, Austin M~E, Callahan K~J, Holcomb C~T, Lestz J~B, Liu L,
  McKee G~R, Schmitz L, Van~Zeeland M~A and Yan Z 2026 {\em Phys. Rev. Lett.\/}
  {\bf 137}(9) 095101

\bibitem{LChenNF2025}
Chen L, Liu P, Ma R, Lin Z, Qiu Z, Wang W and Zonca F 2025 {\em Nuclear
  Fusion\/} {\bf 65} 016018

\bibitem{ZQiuPST2025b}
QIU Z, HU Y, SHEN W, BAO J, REN Z, ZHENG Y, SU P, XU Y, WANG F, CHEN J and LI G
  2025 {\em Plasma Science and Technology\/} {\bf 27} 104005

\bibitem{LChen2026a}
Chen L and Ma R 2026 Beat-driven of zs by tae in weakly and strongly driven
  limits

\bibitem{EFriemanPoF1982}
Frieman E~A and Chen L 1982 {\em Physics of Fluids\/} {\bf 25} 502--508

\bibitem{RIvannovPPCF2026}
Ivanov R, Biancalani A, Bottino A, Gossard D, Hayward-Schneider T, Mishchenko A
  and Wu R 2026 {\em Plasma Physics and Controlled Fusion\/} {\bf 68} 095033

\bibitem{CHsuNF2026}
Hsu C~C, Chen S~H and Chen L 2026 {\em Nuclear Fusion\/} {\bf 66} 036019

\end{thebibliography}
\end{document}